\documentclass[aps,prb,twocolumn,superscriptaddress,floatfix,hyperref=pdftex]{revtex4-1}
\usepackage{times}
\usepackage{graphicx}
\usepackage{dcolumn}
\usepackage{bm}
\usepackage{color}
\usepackage{amsmath}
\usepackage{amssymb}
\usepackage{braket}
\newsavebox{\mybox}

\usepackage[colorlinks=true]{hyperref}

\makeatother

\begin{document}

\title{Fractionalizing a local pair density wave: a good ``recipe'' for
opening a pseudo-gap}

\author{M. Grandadam}

\affiliation{Institut de Physique Th\'eorique, Universit\'e Paris-Saclay, CEA, CNRS, F-91191 Gif-sur-Yvette, France.}

\author{D. Chakraborty}

\affiliation{Institut de Physique Th\'eorique, Universit\'e Paris-Saclay, CEA, CNRS, F-91191 Gif-sur-Yvette, France.}

\author{C. P\'epin}

\affiliation{Institut de Physique Th\'eorique, Universit\'e Paris-Saclay, CEA, CNRS, F-91191 Gif-sur-Yvette, France.}

\begin{abstract}
We give a concise version of a recently proposed concept of fractionalization of an order parameter, thus generating a constraint through a fictitious gauge field \cite{Chakraborty19}. We argue that this
new line of approach is key to explain the longstanding mystery of the pseudo-gap phase in cuprate superconductors. For example, the fractionalization of a finite momentum, charge two state living on latice bonds- also
called Pair Density Wave, into a particle-particle and a particle-hole pair leads to the opening of a gap in the fermionic spectrum. It induces ``phase-locking'' between the particle-particle and particle-hole pairs. We describe the formation of the Fermi arcs in the spectrum and give an account of recent Raman spectroscopy results from a minimal microscopic model. We relate the ``phase-locking'' to intriguing STM experimental observations. 
\end{abstract}
\maketitle

\section{Introduction}

The field of cuprate superconductors is more than thirty years old, but still a few mysteries remain to be understood. One is the origin of the pseudo-gap (PG) phase which occurs above the superconducting state (SC), in the underdoped region of the phase diagram\cite{Alloul89,Warren89}. Despite a huge body of theoretical and experimental investigation, there is still no consensus about explaining this mysterious phase.\\ 
The theoretical approaches can be cast into three broad categories. In the first one, the phase diagram of the cuprates is seen as being dominated by the presence of a Mott insulator at half-filling. This leads to very strong correlations between electrons, so strong that the electron ``fractionalizes''\cite{Coleman84,Baskaran88,Nagaosa90,Lee92,Senthil:2000eb} between more elementary parts, most often a spinon- the elementary excitation carrying spin, and a holon- the elementary excitation carrying charge. This fractionalization leads to the gapping out of the Anti-Nodal (AN) region of the Brillouin zone and to a pre-cursor phase with unusual magnetic properties, for example a spin liquid phase, like in the Resonant Valence Bond (RVB)\cite{Anderson87} proposal. In the second approach, the under-doped region is seen from the ``right-hand side'' within oxygen doping. Above the doping $p\sim20\%$ the system is a Landau Fermi liquid at zero temperature, which means that free quasi-particles can be seen as the elementary excitations. Not only the electron is not fractionalized between more elementary parts, but interactions are moderate, leading to a valid realization of the Fermi liquid phenomenology. From this side, the underdoped phase of the cuprates looks very much like a hidden order phase - a phase whose order we don't yet know, which is terminating around $p\sim18\%$ with a Quantum Critical Point (QCP) having thermodynamic singularities. The third viewpoint is to consider that the under-doped regime is dominated by a huge amount of fluctuations. The approach to localization- Mott transition here, is known to provide wild fluctuations of the phase of the electrons, hence of all the fields present in the system\cite{EmeryVJ:1995dr}. Fluctuations of the Cooper pairs, or for some theories, of ``pre-formed'' pairs\cite{Norman:1998va,Chien2009}, have naturally more strength in the nodal part of the Fermi surface than in the AN one, since the gap vanishes at the four nodal points. This is the common explanation for the opening of Fermi arcs above the superconducting temperature $T_{c}$ in the nodal region of the Brillouin zone.\\
The long-standing observation through ARPES, of such ``Fermi arc''\cite{} which breaks the Luttinger sum rule, is one of the first experimental support for a theory based on phase fluctuations. Later, it was observed, through ARPES again, that in the AN region, the gap behaves in a strange way, not following the ordering SC temperature $T_{c}$ with doping, but rather following the PG temperature $T^{*}$. Raman spectroscopy in the $B_{1g}$ channel\cite{LeTacon06} - which is probing preferentially the AN region of the Brillouin Zone (BZ), shows a pair breaking peak below $T_{c}$, at energy scales of the same order as the gap observed in ARPES, but which again follows $T^{*}$ rather than $T_{c}$ with doping. This set of data was very supportive of the preformed-pairs scenario. Since a precursor was observed in the particle-particle (PP) channel, it was natural to infer that at a higher energy scale $T^{*}$, PP pairs are forming, whereas at a lower energy scale $T_{c}$, phase coherence sets in. This strong phenomenology was put experimentally to the test through transport measurements\cite{RullierAlbenque:2011ji}, SC scanning microscopy\cite{Bergeal:2008gf} or careful study of the Nernst effect\cite{Li:2010gj,CyrChoiniere:2018ed}. All this experimental evidence concurred to show that PP phase fluctuations are only present up to an intermediary energy scale $T_{c}'$ which is far below $T^{*}$.\\\\ 
Recent Raman experiments\cite{Loret19} have brought an important new element in this discussion. Data in the $B_{2g}$ channel for a series of under-doped cuprates ( which is probing preferentially the nodal region, as pictured in Fig.\ref{fig:gapeq}) has shown for the first time a pair breaking peak in the particle-hole (PH) channel. This peak is observed below the temperature $T_{co}$, attributed by many experimental probes like X-Ray\cite{Chang12,Blanco-Canosa13,Campi15,Blackburn13a,Ghiringhelli12,Gerber:2015gx,Chang16} and NMR measurements\cite{Wu11,Wu13a,Wu:2015bt} to the ordering of incommensurate modulations in the charge channel. Remarkably, the pair breaking gap observed in the PH channel doesn't follow $T_{co}$ but $T^{*}$. The situation thus becomes symmetric between the PP and PH channels. Both have a precursor to the ordered phase, whose gap is following the $T^{*}$ line with doping. Both gaps are of the same order of magnitude and are related to the formation of the PG. This recent observation led us to re-consider the preformed pairs scenario, but this time not with preformed pairs only in the PP channel but with an entanglement of both the PP and PH channels.

\section{The main idea}

In order to account for the symmetric features observed both in the PP and PH channel, we make the following Ansatz, that the PG state is a coherent superposition of PP and modulated PH pairs, with the
wave function 
\begin{align}
\ket{PG} & =\hat{\chi}_{ij}\ket{0}+\hat{\Delta}_{ij}\ket{0},\label{eq:1}
\end{align}

where $\ket{0}$ denotes the quantum vacuum of the electronic system, and the operators $\hat{\chi}_{ij}=\sum_{\sigma}\hat{d}c_{i\sigma}^{\dagger}c_{j\sigma}e^{i\mathbf{Q}\cdot\mathbf{r}_{ij}+i\theta_{\chi}}$ and $\hat{\Delta}_{ij}=\sum_{\sigma}\hat{d}\sigma c_{i\sigma}c_{j-\sigma}e^{i\theta_{\Delta}}$ are respectively the operators for PH modulations and PP pairs on a bond $\left\langle i,j\right\rangle $. Within the Ansatz Eq.(\ref{eq:1}), PP and PH pairs can be thought as being entangled in a coherent superposition of ``dead cat'' and ``alive cat'' in the Schr\"odinger's thought experiment. The probability $\braket{PG|PG}=\left(E^{*}\right)^{2}$ sets up an energy scale  
\begin{align}
\sqrt{\left|\chi_{ij}\right|^{2}+\left|\Delta_{ij}\right|^{2}} & =E^{*},\label{eq:2}
\end{align}
which in this theory is identified with the PG scale of cuprate superconductors. Before showing how the PG scale $E^{*}$ manifests itself in one particle and two-particles probe experiments, we give two equivalent theoretical perspectives to understand the Ansatz Eq.(\ref{eq:1}).

First, Eqs.(\ref{eq:1}) and (\ref{eq:2}) taken together can be understood as the fractionalizaton of a bond PDW $\Delta_{PDW}$ into two elementary parts elementary part, PP ($\Delta_{ij}$) and PH ($\chi_{ij}$) preformed pairs. The fractionalization Ansatz goes as  
\begin{align}
\Delta_{PDW} & =\Delta_{ij}\chi_{ij}^{*},\label{eq:3}\\
\mbox{with } & \Delta_{ij}^{*}\Delta_{ij}+\chi_{ij}^{*}\chi_{ij}=1,\nonumber 
\end{align}
the second line in Eq.(\ref{eq:3}) being equivalent to the constraint Eq.(\ref{eq:2}) and the subscripts $\left(i,j\right)$ denote nearest neighbor bonds. The analogy with U(1) gauge theories where the electron
is fractionalized is straightforward. Take for example the slave-boson formulation where the electron creation operator $c_{i\sigma}^{\dagger}$ is fractionalized into a neutral ``spinon'' $f_{i\sigma}^{\dagger}$ and a charged ``holon'' $b_{i}$. We have  
\begin{align}
c_{i\sigma}^{\dagger} & =f_{i\sigma}^{\dagger}b_{i},\label{eq:4}\\
\mbox{with } & \sum_{\sigma}f_{i\sigma}^{\dagger}f_{i\sigma}+b_{i}^{\dagger}b_{i}=1.\nonumber 
\end{align}
This rewriting introduce a new phase degree of freedom as the electron creation operator stay invariant under the local transformation
\begin{align}
f_{i\sigma}\rightarrow e^{i\theta}f_{i\sigma}, &  & b_{i}\rightarrow e^{i\theta}b_{i}.\label{eq:5}
\end{align}\
The corresponding U(1) gauge field associated to this transformation, $\alpha_{\mu}=\partial_{\mu}\theta$, will be related to the constraint in Eq.\eqref{eq:4}.\\

Note that with respect to an external electro-magnetic (EM) field $A_{\mu}$, $b_{i}$ is charged whereas $f_{i\sigma}$ is neutral. Similarly, within the fractionalization of a PDW Eq.(\ref{eq:3}), the field $\Delta_{ij}$ carries EM charge-2 whereas the field $\chi_{ij}$ is, to first approximation, neutral. Our corresponding U(1) gauge theory thus has the ``fictitious'' gauge field $\alpha_{\mu}$ corresponding to the invariance with respect to the local transformation of the global phase of the spinor: 
\begin{align}
\Psi_{ij} &=\left(\Delta_{ij},\chi_{ij}\right)^{T}, \nonumber \\
\Delta_{ij} &\rightarrow e^{i\theta}\Delta_{ij}, \qquad \chi_{ij}\rightarrow e^{i\theta}\chi_{ij}.\label{eq:6}
\end{align}

\section{Effective Field theories}

In this section, we describe two ways to derive the effective field theory for
the pseudogap state. The constraint Eq.(\ref{eq:2}) will be
related to the freezing of the global phase of the spinor $\Psi_{ij}$. 

\subsection{The chiral model and its mapping to the $CP^{1}$ model}\label{chiral model}

The first convenient way to express the fractionalization of the PDW order is to start form a chiral $SU(2)$ model for a fluctuating PDW and to use a mapping onto a $CP^1$ model\cite{Perelomov81}. We start with an action of the form :
\begin{equation}
\mathcal{S}_{SU(2)} = \int d^dx\ \partial_{\mu}\Delta_{PDW} \partial_{\mu}\Delta_{PDW}^*,\label{eq:7}
\end{equation}
where we omitted a possible potential term. By writing the PDW field as in Eq.\eqref{eq:3} and using the associated constraint, the action Eq.\eqref{eq:7} can be rewritten as 
\begin{align}
\mathcal{S} =& \int d^{d}x ~ \left[ |\partial_{\mu}\Delta_{ij}|^2 + |\partial_{\mu}\chi_{ij}|^2 \right. \nonumber\\
&\left. -\left( \Delta_{ij} \partial_{\mu}\Delta_{ij}^{*}-\chi_{ij}^* \partial_{\mu}\chi_{ij} \right)\left( \Delta_{ij}^* \partial_{\mu}\Delta_{ij}-\chi_{ij} \partial_{\mu}\chi_{ij}^* \right) \right].\label{eq:8}
\end{align}
We now note that under the phase transformation $\Delta_{ij} \rightarrow \Delta_{ij} e^{i\theta}$ and $\chi_{ij} \rightarrow \chi_{ij} e^{i\theta}$, which leave our original action Eq.\eqref{eq:7} unchanged, the second term of Eq.\eqref{eq:8} then transform as
\begin{align}
& \left( \Delta_{ij} \partial_{\mu}\Delta_{ij}^{*}-\chi_{ij}^* \partial_{\mu}\chi_{ij} \right) \nonumber \\
& \rightarrow \left( \Delta_{ij} \partial_{\mu}\Delta_{ij}^{*}-\chi_{ij}^* \partial_{\mu}\chi_{ij} \right)-i\partial_{\mu}\theta. \label{eq:9}
\end{align}
Thus we can identify it with a gauge field and rewrite the action Eq.\eqref{eq:8} as
\begin{align}
\mathcal{S}_{CP^1} =& \int d^{d}x ~ |\partial_{\mu}\Psi_{ij}|^2 + \alpha_{\mu} \bar{\alpha}_{\mu} \nonumber \\
=& \int d^{d}x ~ |D_{\mu}\Psi_{ij}|^2, \label{eq:10}
\end{align}
where we used the spinor in Eq.\eqref{eq:6} and we have introduce the covariant derivative $D_{\mu} = \partial_{\mu}-i\alpha_{\mu}$. This action is the one for a $CP^{1}$ model and is invariant under the joint transformation
\begin{equation}
\Psi_{ij} \rightarrow e^{i\theta}\Psi_{ij}, \qquad \alpha_{\mu} \rightarrow \alpha_{\mu}-\partial_{\mu} \theta, \label{eq:11}
\end{equation}
reflecting the fact that this phase transformation is introduced by our decomposition Eq.\eqref{eq:3} and has to leave our original model Eq.\eqref{eq:7} invariant. Note that in this case, the value of the gauge field $\alpha_{\mu}$ is fixed and given by $\alpha_{\mu} = -i\left(\Delta_{ij}\partial_{\mu}\Delta_{ij}^*-\chi_{ij}^*\partial_{\mu}\chi_{ij}\right)$. This is equivalent to considering the global phase of the spinor $\Psi_{ij}$ as being fixed. Furthermore, the two fields $\Delta_{ij}$ and $\chi_{ij}$ are linked by the constraint of Eq.\eqref{eq:3}. \\
This constraint does not however imply that the PP or PH pairs condense. In fact, just below the pseudogap temperature the amplitude of both $\Delta_{ij}$ and $\chi_{ij}$ are still fluctuating. It is only at lower temperatures that these amplitudes become finite. In our case we can argue on phenomenological ground that the PH pairs will condense first at a temperature $T_{co}$, where charge modulation is observed experimentally. In this case it becomes possible to observe the PDW order as it is reconfined. This order is however not a long range order as we still have fluctuations of the relative phase between the two components of $\Psi_{ij}$. This remaining phase is actually linked to the coupling with an EM field as it will be more apparent in the following section. The freezing of this phase will lead to the superconducting transition at $T_c$ (see Fig.\ref{fig:scdiagram}).

\subsection{The $U\left(1\right)\times U\left(1\right)$ theory}\label{u(1)xu(1)}
The same theory can be obtained by starting from a general action for two complex fields with their respecting gauge degree of freedom,
\begin{align}
\mathcal{S} &= \int d^dx\ |D_{\mu}z_1|^2 + |\tilde{D}_{\mu}z_2|^2,\nonumber \\
D_{\mu} &= \partial_{\mu}-if_{\mu}, \quad \tilde{D}_{\mu} = \partial_{\mu}-i\tilde{f}_{\mu}. \label{eq:12}
\end{align}
If we identify the two complex field with our previous order parameters $z_1 = \Delta_{ij}$ and $z_2=\chi_{ij}$ and their respective gauge degree of freedom $f_{\mu} = 2A_{\mu}+\alpha_{\mu}$ and $\tilde{f}_{\mu} = \alpha_{\mu}$ we can rewrite this action with the spinor
\begin{equation}
\Psi_{ij}^{\dagger} = \left(|\Delta_{ij}|e^{i\theta_{\Delta}}, |\chi_{ij}|e^{i\theta_{\chi}}\right) = e^{i\theta}e^{i\tau_z\phi}\left(|\Delta_{ij}|,|\chi_{ij}|\right), \label{eq:13}
\end{equation}
where we have made apparent the global ($\theta$) and the relative ($\phi$) phase between the two component of the spinor and used the third Pauli matrix $\tau_z$. We thus end up with an action,
\begin{align}
\mathcal{S} &= \int d^dx |D_{\mu}\Psi_{ij}|^2 + V\left(\Psi_{ij},\Psi_{ij}^{\dagger}\right), \nonumber \\
D_{\mu} &= \partial_{\mu}-ia_{\mu}-i\tau_zb_{\mu}. \label{eq:14}
\end{align}
where the gauge fields refer this time to the global and relative phases of the spinor and are associated to the following transformations
\begin{align}
\Psi_{ij} \rightarrow e^{i\theta}\Psi_{ij} \qquad a_{\mu}\rightarrow a_{\mu}+i\partial_{\mu} \theta \nonumber, \\
\Psi_{ij} \rightarrow e^{i\tau_z\phi}\Psi_{ij} \qquad b_{\mu}\rightarrow b_{\mu}+i\partial_{\mu} \phi. \label{eq:15}
\end{align}
We can also directly relate this new gauge fields to the previous ones
\begin{equation}
a_{\mu} = A_{\mu} + \alpha_{\mu}, \qquad b_{\mu} = A_{\mu} \label{eq:16}.
\end{equation}
The action Eq.\eqref{eq:14} is similar to the one we obtained in Eq.\eqref{eq:8} when we started from a model for a fluctuating PDW. However we didn't impose any constraint between the two fields $\Delta_{ij}$ and $\chi_{ij}$ in Eq.\eqref{eq:14}, this result in the two gauge fields fluctuating.\\
It is now possible to imagine that a Higgs mechanism will take place at $T^*$, freezing one of the phase and giving a mass to the corresponding gauge field. Applying this idea to the global phase $\theta$ will lead to a mass for the $a_{\mu}$ gauge field equal to the spinor amplitude, i.e 
\begin{equation}
m_{a} = |\Psi_{ij}|^2 = \sqrt{|\Delta_{ij}|^2 + |\chi_{ij}|^2} := E^*. \label{eq:17}
\end{equation}

\begin{figure}[t]
\includegraphics[width=0.9\linewidth]{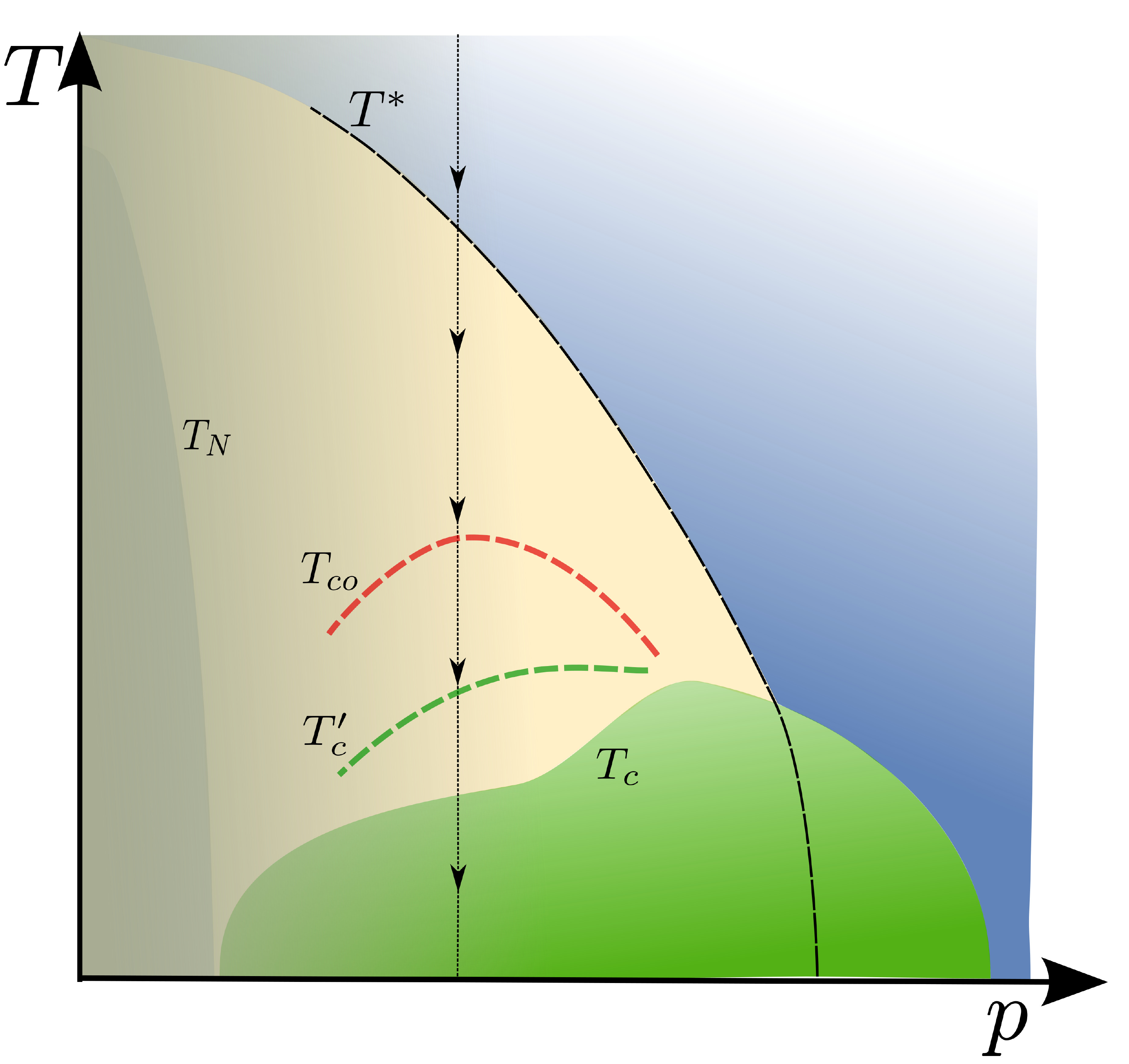}
\caption{Schematic temperature (T)- hole doping ($p$) phase diagram\cite{Chakraborty19} for a cuprate superconductor. As we decrease temperature (black line), we have a first phase transition at $T^*$. This sets the pseudogap energy scale and induces a constraint between the two orders as shown in Eq.\eqref{eq:17}. At lower temperatures $T_{co}$ and $T^{\prime}_c$ the amplitudes of the two orders become finite. These are not phase transition but crossover lines. A second Higgs mechanism occurs at $T_c$, where the relative phase of the spinor gets frozen.}
\label{fig:scdiagram} 
\end{figure}

Hence, we recover a constraint by minimizing the action with respect to a gauge field. If we identify the temperature at which the Higgs mechanism describe in this section occurs with the transition at which the PDW field get fractionalized, these two theories are equivalent. In both cases, we end up with a theory of two amplitudes $|\Delta_{ij}|$, $|\chi_{ij}|$, linked by a constraint, and one phase $\phi$ fluctuating. \\
Identifying the general setup presented here with cuprate superconductors we can describe the phase diagram in the underdoped regime when we lower the temperature as depicted in Fig.\ref{fig:scdiagram}. A first phase transition occurs at $T^*$ when the global phase $\theta$ gets frozen and the constraint between the particle-particle pair field $\Delta_{ij}$ and the particle-hole field $\chi_{ij}$ sets an energy scale $E^*$. The finite amplitude for the spinor $\Psi_{ij}$ at $T^*$ does not however mean a condensation of $|\Delta_{ij}|$ or $|\chi_{ij}|$ as the constraint in Eq.\eqref{eq:2} can be satisfied by the fluctuations of both fields. We can thus think that these fields will acquire a finite amplitude at lower temperatures $T_{co}$ and $T_c^{\prime}$ for the PH and PP pairs respectively. In this sense this is similar to a preformed pairs description of the pseudogap phase but where two kinds of pairs, particle-particle and particle-hole, are entangled. Lastly, the remaining relative phase $\phi$ will get frozen at a lower temperature $T_c$ which is then identified with the superconducting transition as the EM gauge field gets massive, leading to Meissner effect.

\section{Microscopic model : Raman, ARPES and phase coherence in vortices}

We now present a minimal microscopic model that allows us to extract some relevant experimental result using mean-field theory. \\
We start by considering electrons hopping on a square lattice and interacting through short-range antiferromagnetic spin fluctuations. If this is known to give rise to both superconductivity and charge density wave (CDW), we here add an off-site density-density interaction to favour the CDW state at the mean-field level. Our starting real space Hamiltonian thus reads:
\begin{equation}
H = \sum_{ij} -t_{ij} \left(c_{i \sigma}^{\dagger}c_{j\sigma}+h.c. \right)+J_{ij} \bm{S}_i\ \bm{S}_j + V_{ij}n_i n_j, \label{eq:18}
\end{equation}
where $c^{\dagger}_{i \sigma}$ ($c_{i,\sigma}$) creates (destroys) an electron at the site i on a square lattice and we can write the spin operator $S^a_i = c^{\dagger}_{i \sigma}\tau^a_{\sigma \sigma'}c_{i \sigma'}$ or the number operator $n_i = c^{\dagger}_{i,\sigma}c_{i,\sigma}$ with appropriate spin summation. The hopping parameter $t_{ij}$ is taken up to the 4$^{th}$ nearest-neighbours and taken to reproduce the band dispersion of Hg-1201\cite{Loret19}. $J_{ij}$ denotes the antiferromagnetic interaction and is positive while $V_{ij}$ is the off-site density interaction.
We then go to momentum space and decouple the interaction terms with a particle-particle pairing field and a particle-hole pairing field at finite momentum defined by
\begin{align}
\Delta_k & = \sum_{\sigma} \sigma c_{k\sigma}c_{-k -\sigma}, \nonumber \\
\chi^Q_k &= \sum_{\sigma} c^{\dagger}_{k \sigma}c_{k+Q \sigma} .\label{eq:19}
\end{align}
This leads us, after going in a path integral formalism, to the action
\begin{align}
\mathcal{S}_{micro} &= \ \mathcal{S}_{c,0} + \mathcal{S}_{c,\Delta,\chi}+\mathcal{S}_{\Delta,\chi,0} \nonumber ,\\
\mathcal{S}_{c,0} &= \sum_{k,\epsilon} c_{k \sigma}\left(\epsilon-\xi_k \right)c^{\dagger}_{k \sigma} \nonumber, \\
\mathcal{S}_{c,\Delta,\chi} &= \sum_{k,q} \Delta^*_k c_{k+q \sigma}c_{-k-q -\sigma} + \chi^{Q*}_{k}c^{\dagger}_{k+q \sigma}c_{k+q+Q \sigma} \nonumber \\
 &+ \Delta_{k+q} c^{\dagger}_{-k -\sigma}c^{\dagger}_{k \sigma} + \chi^{Q}_{k+q}c^{\dagger}_{k+Q \sigma}c_{k \sigma} \nonumber, \\
\mathcal{S}_{\Delta,\chi,0} &= \sum_{k,q} \frac{\Delta^*_k \Delta_{k+q}}{J_-\left(q\right)} + \frac{\chi^{Q*}_k \chi^Q_{k+q}}{J_+\left(q\right)}, \label{eq:20}
\end{align}
where we used $J_{\pm}\left(q\right) = 3J\left(q\right) \pm V\left(q\right)$ and $\xi_k$ is the electronic dispersion. The exact form of $J_{\pm}\left(q\right)$ is not important and we only take it to be peaked around $\left( \pi,\pi \right)$ with a broadening $\kappa_{AF}$ taken form Inelastic Neutron Scattering\cite{Chan16}. From here, we can obtain the self-consistent gap equation for both $\Delta_k$ and $\chi_k$ in a standard way, by integrating the fermionic degrees of freedom and minimizing the action with respect to the relevant field:
\begin{align}
\Delta_k =& -T \sum_{q,\omega} J_-\left(q\right)\frac{\Delta_{k+q}}{G^{-1}_{0,k+q,\omega+\epsilon}G^{-1}_{0,-k-q,\omega+\epsilon}-|\Delta_{k+q}|^2}, \nonumber \\
\chi^Q_k =& -T \sum_{q,\omega} J_+\left(q\right)\frac{\chi^Q_{k+q}}{G^{-1}_{0,k+q,\omega+\epsilon}G^{-1}_{0,k+q+Q,\omega+\epsilon}-|\chi^Q_{k+q}|^2}, \label{eq:21}
\end{align}
where $\omega$ ($\epsilon$) is a bosonic (fermionic) imaginary Matsubara frequency, and $G_{0,k,\epsilon}=\left(\epsilon - \xi_k \right)^{-1}$ is the free electronic Green's function.
Solving these gap equations allows us to see how can the two order coexist in momentum space. Indeed, if the previous constraint Eq.\eqref{eq:2} is local and does not favor coexistence in real space, it allows for the two orders to live in different regions of the Brillouin zone. We thus consider that only one gap will open at each k-point, the one with the largest amplitude. Here the principal driving mechanism for pairing comes from the antiferromagnetic fluctuations and we take the strong electronic correlations into account only at the mean-field level by renormalizing the hopping and the interaction via Gutzwiller parameters \cite{Anderson:2004cz}. \\

\begin{figure}
\includegraphics[width=1.0\linewidth]{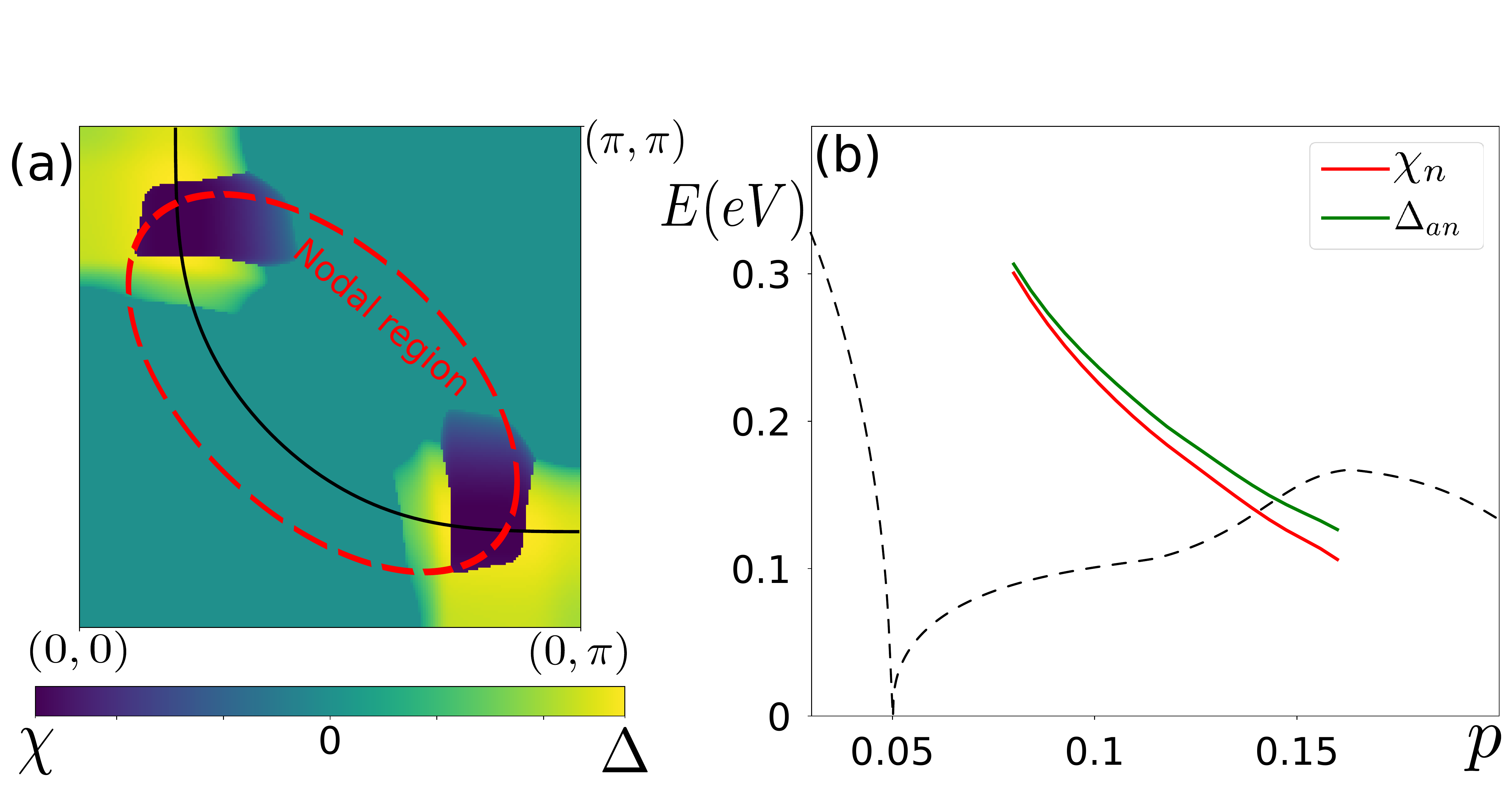}
\caption{Results\cite{Chakraborty19} from the gap equations Eq.\eqref{eq:21} \textbf{(a)} Superconducting ($\Delta$) and the Charge order with axial wave-vector ($\chi$) gaps in the first quadrant of the Brillouin zone for doping $p=0.12$. At each k-point only the largest of the two gap is shown (see text). The black line indicates the non-interacting Fermi surface and the red dotted line indicate the nodal regions probed in $B_{2g}$ Raman response.  While the SC gap opens in the AN region, the CDW gap prevails in the nodal region. \textbf{(b)} The doping dependence of the SC gap averaged in the AN region ($\Delta_{an}$) and the CDW gap averaged in the nodal region ($\chi_{n}$). They both behave similarly as a function of doping in the range $0.08 >p>0.16$ with $\Delta_{an} \approx \chi_{n}$. This result fits the experimental trends\cite{Loret19} obtained in Raman spectroscopy very well. Parameters used for this plot are $J=350 \ meV$, $V=J/20$ and $\kappa_{AF} = 0.1\ r.l.u$. The dashed lines schematically indicate the doping region where antiferromagnetic order and the superconducting dome lies.}
\label{fig:gapeq}
\end{figure}

\begin{figure}
\includegraphics[width=1.0\linewidth]{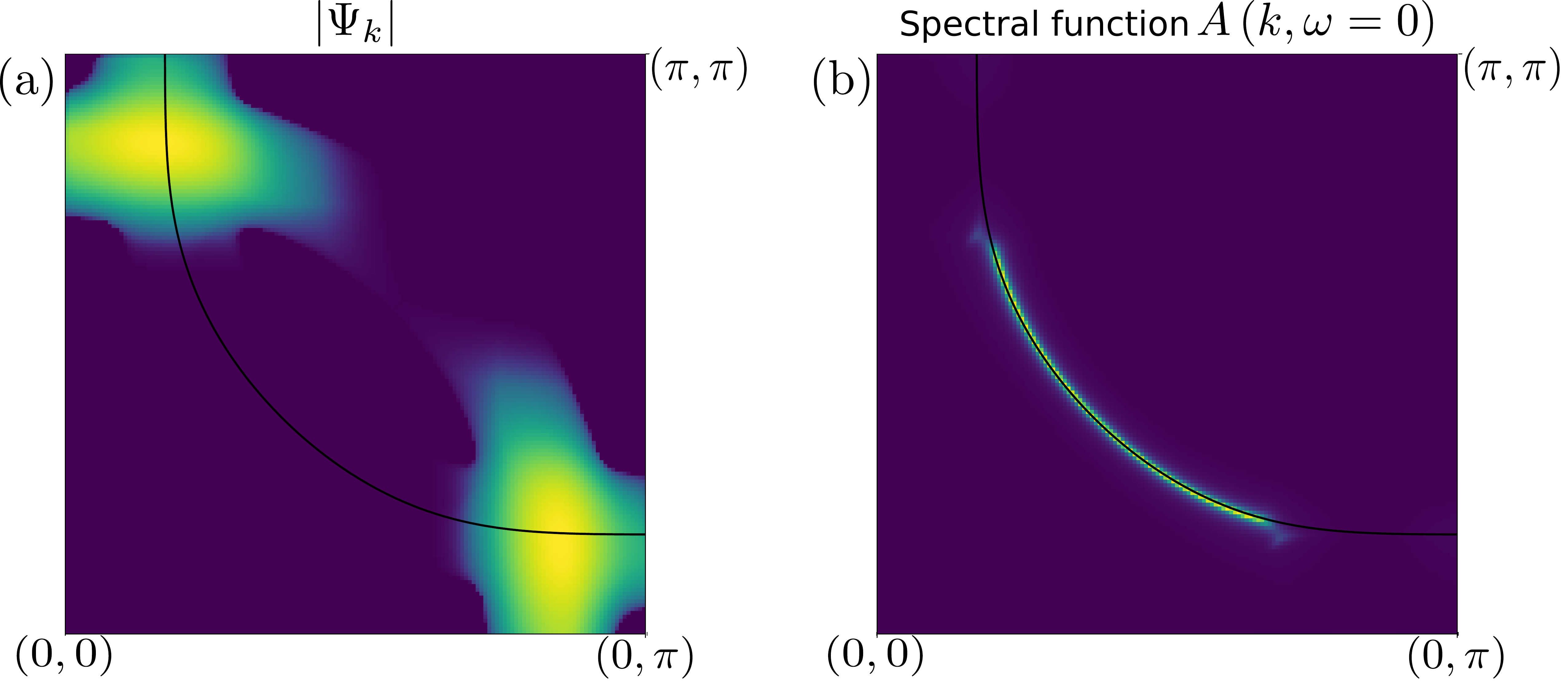}
\caption{ \textbf{(a)} Solution for the spinor amplitude $|\Psi_k|$ from solving the gap equation Eq.\eqref{eq:24} . \textbf{(b)} Electronic spectral function at $\omega = 0$ obtain from the Green function Eq.\eqref{eq:22} with the solution shown in panel \textbf{(a)} for $|\Psi_k|^2$. We see the formation of a Fermi arc as the AN region is gapped out. The black line indicates the non-interacting Fermi surface and parameters are the same as given in Fig.\ref{fig:gapeq}.}
\label{fig:ARPES}
\end{figure}

Superconductivity being a full Fermi surface instability will gap out a larger part of the Brillouin zone than the CDW order. However, owing to the density-density interaction, the latter will prevail when its intrinsic wave-vector links two parts of the Fermi surface. Because of the form of our interaction $J_{\pm}\left(q\right)$, and guided by experimental observations, we take the wave-vector of the CDW order to be axial and connecting hot-spots, points of the Fermi surface related by the AF wave-vector $\left(\pi,\pi\right)$, in the first Brillouin zone. One example of the obtained solutions is shown in Fig.\ref{fig:gapeq} where we see that the CDW is dominant around the hot-spot, where the nesting condition is satisfied, while the SC gaps out the antinodal region. Nevertheless the two orders are close to each other in energy. If we look at the value of the SC gap in the antinodal region and compare it with the maximum of the CDW gap in the nodal region for different doping we can see that they both decrease with increasing doping while remaining almost degenerate. This behaviour is consistent with recent Raman spectroscopy experiments\cite{Loret19} that extracted these two energy scales. They were indeed able to extract the two energy scales separately by probing preferentially the nodal or the antinodal region and found that the SC and CDW gap are of the same order and both decrease monotonously with doping.\\\\
\indent Following the effective theories given in Sec.\ref{chiral model} and Sec.\ref{u(1)xu(1)}, the pseudogap phase is described by the spinor $\Psi_{ij}$ having a finite amplitude but none of the two orders $\Delta_{ij}$ and $\chi_{ij}$ being condensed, i.e. their respective mean-field value is zero. To take this into account we write the electronic Green's function depending on the spinor $\Psi_{k}=\left(\Delta_{k},\chi_{k}\right)^{T}$,
\begin{align}
G_{k,\epsilon}^{-1} =& \epsilon - \xi_k -\frac{|\Delta_k|^2}{\epsilon +\xi_k}-\frac{|\chi^Q_k|^2}{\epsilon -\xi_{k+Q}} \nonumber \\
= & \epsilon - \xi_k -\frac{|\Psi_k|^2}{2}\left(\frac{1}{\epsilon +\xi_k}+\frac{1}{\epsilon -\xi_{k+Q}}\right) \nonumber \\
= & \epsilon - \xi_k -|\Psi_k|^2\tilde{G}_{k,\epsilon}. \label{eq:22}
\end{align}
The same rewriting can be made for the third term in Eq.\eqref{eq:20} and leads to our action in the pseudogap phase,
\begin{equation}
\mathcal{S}_{PG} =  \sum_{k,q,\epsilon} \frac{\Psi_k \Psi^{\dagger}_{k+q}}{\tilde{J}\left(q\right)}-Tr\ ln\left(G_{0,k,\epsilon}^{-1}-|\Psi_k|^2\tilde{G}_{k,\epsilon}\right), \label{eq:23}
\end{equation}
with $\frac{1}{\tilde{J}\left(q\right)} = \frac{1}{2}\left(\frac{1}{J_+\left(q\right)}+\frac{1}{J_-\left(q\right)}\right)$.  We can now obtain a mean-field equation determining the amplitude of the spinor by minimizing this action with respect to $\Psi_k^{\dagger}$:
\begin{equation}
|\Psi_k| = -T\sum_{q,\omega} \tilde{J_q}\frac{|\Psi_{k+q}|}{G^{-1}_{0,k+q,\omega+\epsilon}\tilde{G}^{-1}_{k+q,\omega+\epsilon}-|\Psi_{k+q}|^2}. \label{eq:24}
\end{equation}
Solving this gap equation for the same parameters as the ones given in Fig.\ref{fig:gapeq} gives a solution for $|\Psi_k|$ shown in Fig.\ref{fig:ARPES}(a). We see once again that the nesting condition imposed by the modulation vector $Q$ lead to the nodal region being unaffected. The resulting electronic spectral function obtained from the Green function Eq.\eqref{eq:22} and the solution to the gap equation Eq.\eqref{eq:24} is shown in Fig.\ref{fig:ARPES}(b) where we see the presence of Fermi arcs characteristic of the pseudogap phase.\\

\begin{figure}[t]
\includegraphics[width=1.0\columnwidth]{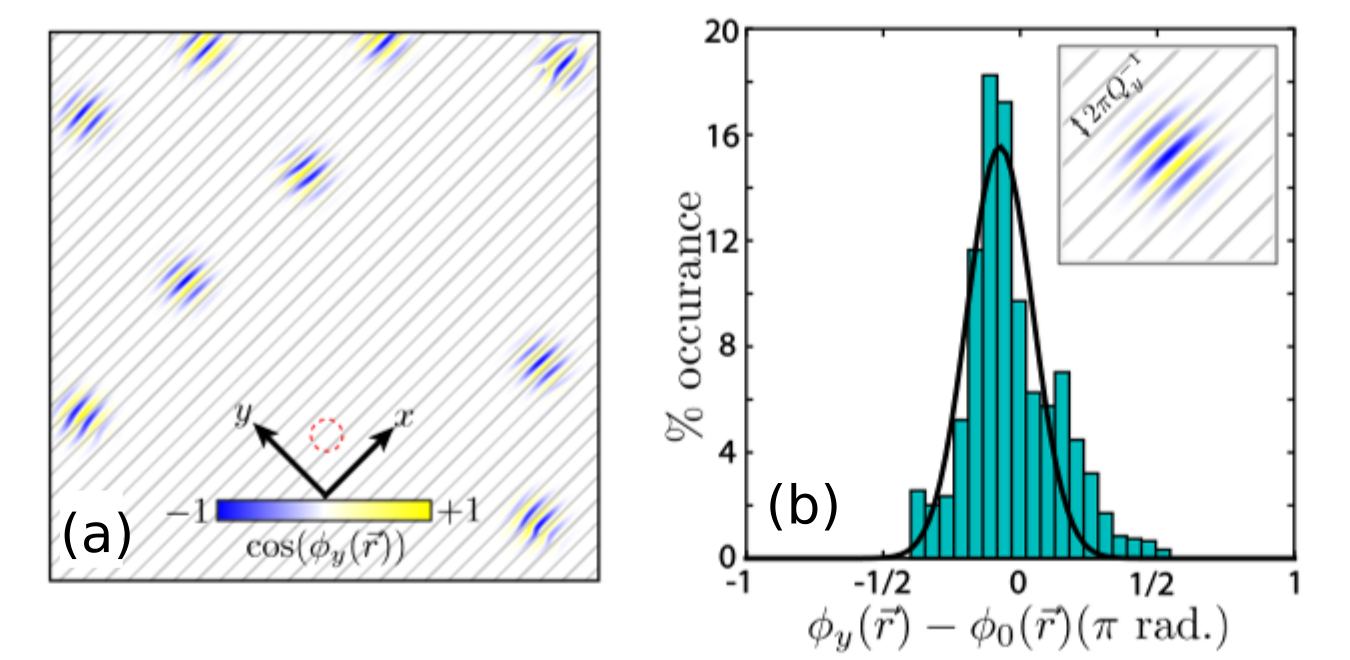} 
\caption{\label{fig:Global-spatial-Charge} Experimental STM data \cite{Hamidian15} showing long-range phase coherence of the CDW order inside SC vortices. \textbf{(a)} The amplitude and phase of the CDM with predominant wave vector $Q_{y}$ inside stable vortex cores. Typical size of a vortex core is of 2 nm shown by the red dotted circle. the phase is encoded by the blue/yellow colours and is measured with respect to the reference phase due to the modulation (shown by the grey lines). \textbf{(b)} Histogram plot of the measured phase of the CDW order with respect to the reference phase. The plot shows that the relative variation of phase is centered around a single value for multiple vortices with a standard deviation of $12\%(2\pi)$. For this plot, 6-9 vortices were used because of the stability issue of vortex\cite{Hamidian15}. }
\end{figure}

Another peculiar experimental observation that can be explained within our approach is the long-range phase coherence of CDW order in SC vortices. Indeed, STM experiments have been able to measure the phase of the CDW order\cite{Hamidian15} that is observed in the vortex core below $T_c$. One of these measurements is presented in Fig.\ref{fig:Global-spatial-Charge}(a) where we see the phase of the charge modulation with respect to the underlying lattice. Even though the CDW order is short-range and the vortices do not overlap, we can see a correlation of the phase between different vortices represented in Fig.\ref{fig:Global-spatial-Charge}(b). While the observation of CDW order in the core of the vortices can be explained by considering competition between the two orders, this long-range phase coherence is more puzzling. We can address this observation by using the phase-locking mechanism discussed in Sec.\ref{u(1)xu(1)}. Indeed, even when the CDW is short-range, the long-range coherence of the SC order will lock the phase of the density modulation in different vortices.\\

\section{Discussion}

We have presented here a new idea to describe the pseudogap phase of underdoped cuprates based on a fractionalized Pair Density Wave. The PDW field is written as a product of a uniform particle-particle pairing field (SC) and a modulated particle-hole pairing field (CDW) with wave-vector $Q$ which are constrained as given in Eq.\eqref{eq:3}. This is analogous to the fractionalization of electrons which can be decomposed into spinons and holons linked by a constraint. In our case, the constraint is directly related to the pseudogap and sets its energy scale. The pseudogap is then a quantum superposition of SC and CDW (see Eq.\eqref{eq:1}). Remarkably, we obtained the same theory for the pseudogap by considering a Higgs mechanism at $T^*$ as shown in Sec.\ref{u(1)xu(1)}. This Higgs mechanism results in the phases of the two orders being locked in the pseudogap phase. We showed that there are two ways to derive the effective theory below $T^*$ and that a simple microscopic model can reproduce several experimental observations. In particular, we obtain the formation of Fermi arcs above $T_c$ where the electronic spectral function gets modified due to the finite amplitude of the spinor $|\Psi|$. We also make a connection with Raman spectroscopy where the amplitudes of the PP and PH gaps are shown to be very close.\\

We now compare our formalism with other recent works in the field. The idea that a PDW order is driving the pseudogap phase has been used in different theoretical approaches\cite{Agterberg19}. In these scenarios, the PDW is the parent order\cite{Dai18,Wang18} from which secondary orders, like the CDW order, emerge. In these cases, none of the electron or the PDW order is fractionalized, in strong contrast with our proposition.\\

Previous theories also considered connecting the SC and CDW order through an emergent $SU(2)$ symmetry\cite{Metlitski10b,Efetov13,Kloss:2016hu,Montiel16}. Remarkably, the constraint obtained through the fractionalization of the PDW order is the same as the one implied by the emergent symmetry. We can thus describe the SC and CDW fluctuations by a $O(3)$ non-linear sigma model. We thus expect that the phenomenological consequences of these two approaches to be the same, with for example the description of the B-T phase diagram\cite{Chakraborty18} or the presence of collective modes\cite{Morice18b}. \\

Our proposal of fractionalization of a PDW is thus new and distinct from previous works. It is fundamentally different from theories considering fractionalization of the electron. There is one other recent proposal based on the fractionalization of an order parameter in the context of cuprates. It is however based on the fractionalization of a fluctuating spin density wave\cite{Sachdev19} which leads to a $SU(2)$ gauge theory. The two approaches have in common the phase transition at $T^*$ and the multiplicity of possible orders in the pseudogap phase\cite{Chakraborty19}.\\

\begin{acknowledgements}

We thank J.C.S. Davis, M.H. Hamidian, Y. Sidis, X. Montiel, S. Sarkar and A. Banerjee for valuable discussions and critical reading of our manuscript. This work has received financial support from the ERC, under grant agreement AdG-694651-CHAMPAGNE. The authors also like to thank the IIP (Natal, Brazil), where the inspiration for this work came from and where parts of this work were done, for hospitality.

\end{acknowledgements}

\bibliographystyle{apsrev4-1}
\bibliography{Cuprates}

\end{document}